\documentclass[
aip,
jcp,
 amsmath,amssymb,
 reprint,%
]{revtex4-1}

\usepackage{graphicx}
\usepackage{dcolumn}
\usepackage{bm}
\usepackage{amsmath,amsfonts,amssymb}
\usepackage{soul}

\usepackage{xcolor}
\usepackage{mathtools}
\usepackage{float}
\usepackage{lipsum}

\usepackage[utf8]{inputenc}
\usepackage[T1]{fontenc}
\usepackage{mathptmx}
\usepackage{etoolbox}

\makeatletter
\def\@email#1#2{%
 \endgroup
 \patchcmd{\titleblock@produce}
  {\frontmatter@RRAPformat}
  {\frontmatter@RRAPformat{\produce@RRAP{*#1\href{mailto:#2}{#2}}}\frontmatter@RRAPformat}
  {}{}
}%
\makeatother
\begin{document}

\preprint{AIP/123-QED}

\title[Bonding and the dynamics of glassy network liquids]{Bonding and the dynamics of glassy network liquids}
\author{M. H. Brown}
\affiliation{
Center for Theoretical Biological Physics, Rice University, Houston, Texas 77005, USA
}
\affiliation{ 
Department of Physics and Astronomy, Rice University, Houston, Texas 77005, USA
}

\author{P. G. Wolynes}%
 \email{pwolynes@rice.edu.}

\affiliation{
Center for Theoretical Biological Physics, Rice University, Houston, Texas 77005, USA
}
\affiliation{ 
Department of Physics and Astronomy, Rice University, Houston, Texas 77005, USA
}
\affiliation{
Department of Chemistry, Rice University, Houston, Texas 77005, USA
}

\date{\today}

\begin{abstract}
The Random First Order Transition (RFOT) theory of glasses provides a unified framework for explaining the observed correlations of the kinetic and thermodynamic behaviors of glass-forming liquids having a wide variety of chemical compositions and interactions. The theory also provides a solid starting point for calculating glassy dynamics starting from the microscopic forces. Network liquids, which interact via long-lived, geometrically constraining interactions, such as covalent bonding, have competing energy scales for bond breaking events and for collective particle rearrangement events. In this paper, we show microscopic calculations via the RFOT theory can predict how glassy dynamics depends on the degree of bonding, focusing on mixtures of network-forming particles with non-bonding impurities as in familiar window glass. By introducing soft-core nonbonding interactions, we show the viscosity and fragility of the network liquid model can be computed as a function of composition, temperature, and density or pressure. We find that the fragility in the strong-bond limit depends only on composition and not on the bond breaking energy and describes well corresponding measurements in sodium or potassium silicates. The model predicts that materials with weaker bonds may show a non-monotonic trend in the fragility as a function of composition. 
\end{abstract}

\maketitle
\section{\label{sec:intro}Introduction}
When a liquid is supercooled, motions even at the molecular level slow down dramatically. The timescale for rearranging the molecular units of a liquid increases from a typical value of around $10^{-12}$ s near the melting point to 100 s at the lower laboratory glass transition temperature $T_g$. Below this temperature, the rearrangements become slower than the duration of most reasonable experiments, and the liquid falls out of equilibrium, forming a glass. In order to understand the “glass transition”, one thus must explain the dramatic increase in rearrangement timescales with decreasing temperature.

The Random First Order Transition (RFOT) theory \cite{kirkpatrick_scaling_1989} explains and indeed predicts many of the regularities observed about the glass transition, most notably the correlations between kinetics and thermodynamics. The RFOT theory describes the liquid as a mosaic of glassy states, which evolves dynamically through cooperative rearrangements. These rearrangements are driven by the configurational entropy $S_c$, which measures the number of alternative glassy states that may be accessed at a given temperature. At the Kauzmann temperature $T_K$, the configurational entropy per particle vanishes as $s_c=\Delta c_p\left(1-\frac{T_K}{T}\right)$ where $\Delta c_p$ is the measured drop in heat capacity between the liquid and glass states. Therefore at $T_K$, there is no driving force to reconfigure so the system would appear to be in an ideal glassy state. For most substances, the timescale for rearrangements above $T_g$ seems to diverge upon cooling as $\tau_\alpha=\tau_0\exp\left(\frac{D T_0}{T-T_0}\right)$ where $\tau_0$ is interpreted as the microscopic timescale, which is on the order of ps in most molecular liquids. The well established coincidence of the thermodynamically defined $T_K$ and the kinetically defined $T_0$ for a huge array of substances is remarkable and would be unexpected in other pictures of the glass transition. The parameter $D$ is called a liquid’s “fragility” and varies between materials. This formula, known as the Vogel-Tamman-Fulcher (VTF) equation, works quite well \cite{angell_formation_1995} as a curve fit to the measured viscosities of supercooled liquids. The RFOT theory not only leads naturally to the VTF equation as an asymptotic result but also predicts correlations between the configurational heat capacity $\Delta c_p$ and kinetic properties, namely the fragility $D$ \cite{xia_fragilities_2000}, the non-exponentiality of relaxations $\beta$ \cite{xia_microscopic_2001}, and the nonlinearity of relaxations below the glass transition \cite{lubchenko_theory_2004}. In supercooled liquids, experiments quantitatively agree with the RFOT theory predictions for these finer correlations, as well as for the coincidence of $T_0$ and $T_K$. In addition, many other dynamic effects such as front propagation through facilitation \cite{wisitsorasak_fluctuating_2013}, detailed structure of the aging dynamics \cite{lubchenko_theory_2004}, the strength of glasses \cite{wisitsorasak_strength_2012}, and the size of cooperatively rearranging regions \cite{berthier_direct_2005} are explained through the RFOT theory. For a review of many of these concepts, see the reviews by Lubchenko and Wolynes \cite{lubchenko_theory_2007} and by Ediger et al. \cite{ediger_glass_2021}.

It has become common to categorize a glass former based on its fragility as being either fragile or strong, but in fact this continuum of behaviors is tunable through varying the composition in mixtures. Strong liquids have nearly Arrhenius dynamics and nearly exponential relaxations. The glass transition temperature and fragility of a given glass former depend ultimately on the nature of its microscopic interactions. The pure organic liquid OTP is fragile, with  $T_g=241$ K while the molecular network-forming liquid $\mathrm{SiO_2}$ is strong, with $T_g$=1500 K \cite{bohmer_nonexponential_1993}. While network glasses with a high degree of bonding (like $\mathrm{SiO_2}$) \cite{bohmer_nonexponential_1993} tend to be quite strong, when the degree of bonding is reduced (by introducing a non-bonding impurity such as $\mathrm{Na_2O}$ or $\mathrm{K_2O}$), these network glass mixtures become progressively more fragile and also exhibit a dramatically lower $T_g$ \cite{poole_low-temperature_1949}. These effects of impurities underly the ancient story of the discovery of glass by the Phoenicians \cite{pliny_the_elder_natural_1965} which argued that sand could be melted through the addition of ash. Network glasses were analyzed earlier using the RFOT theory by Hall and Wolynes \cite{hall_microscopic_2003}. They employed the self-consistent phonon method to determine from a microscopic model of a network liquid how the configurational entropy depends on the number of bonds per particle. Their analysis revealed that as a network glass becomes more thoroughly bonded, $\Delta c_p$ becomes lower, and as a result, the RFOT theory predicts that the glass will be stronger, much as seen experimentally.

Nevertheless, since the temperature dependence of the relaxations of SiO$_2$ appears to be almost perfectly Arrhenius, it has often been thought that the temperature dependence might be explained as simply arising from bond breaking events and that rearrangements occur once one or two bonds break. Both Mott and Turnbull \cite{mott_viscosity_1987, de_neufville_effect_1970} discussed such a scenario decades ago, and it has continued to be invoked \cite{webb_silicate_1997, ojovan_topologically_2006}. Such a rearrangement might be an Arrhenius process with an activation energy of the order of the energy cost of breaking a bond. In this paper we develop the Hall-Wolynes analysis further by accounting for the statistical thermodynamics of thermal bond breaking and show that many features of the activation dynamics of network liquids with breakable bonds are in fact explained by the RFOT theory. Our model is a Lennard-Jones liquid where some pairs of particles are considered bonded and interact via a harmonic bonding potential. This is motivated in part by some biological materials, such as the cytoskeleton, where similar models have provided useful insight \cite{wang_tensegrity_2012}. The model presented here also resembles the models of Molinero and Moore for water \cite{molinero_water_2009} and Stillinger and Weber for silicon \cite{stillinger_computer_1985} in that it treats tetrahedral particles as the fundamental units to which interactions are applied. This simplified picture provides useful insight into the basic physics of network liquids, even if it may ignore many details.

\section{Background}
In the RFOT theory \cite{kirkpatrick_scaling_1989}, we think of a liquid as being described in terms of a “global library” \cite{lubchenko_theory_2004} of glassy states. In mean field theory or in infinite dimensions, these glassy states are regions of phase space which are separated from one another by infinitely high free energy barriers. In 3 dimensional supercooled liquids below the crossover temperature to activated dynamics $T_C$, glassy states are instead separated by finite free energy barriers, with barrier heights that vanish at $T_C$. Because these local minima in the free energy landscape are separated by barriers, it is meaningful to define these glassy states on timescales shorter than the relaxation time, after which the liquid has escaped the local minimum. In each glassy state, the individual particles of a supercooled liquid are taken as vibrating around fixed points representing a globally defined aperiodic crystal structure with some particular vibrational amplitudes. At sufficiently low temperatures, neutron scattering \cite{mezei_intermediate_1999} shows us that a supercooled liquid is well described locally by a single glassy structure, but after many vibrational periods, the particles shift by a distance larger than the vibrational amplitude through cooperative motions. After a cooperative rearrangement event, the liquid must again be well described locally by a different aperiodic crystal structure chosen from the global library.

\subsection{Determining the configurational entropy from the self-consistent phonon method}

The typical glassy configuration from the global library has on average the same energy per particle as the liquid state ensemble but being an individual arrangement, has extensively lower entropy. This entropy difference arises from the number of different particle configurations (aperiodic crystal structures) which are sampled by the liquid at a given temperature, which is why the entropy difference is called the configurational entropy $S_c$ (throughout this paper, extensive thermodynamic quantities are written with capital letters and the corresponding quantities ``per particle'' are written with lowercase letters). The configurational entropy provides the driving force of the dynamics of supercooled liquids and is thus central to the behavior of glassy materials. As long as the configurational entropy is positive, each individual glassy state sampled at that temperature is ultimately metastable to the liquid, which is an ensemble of such structures.

The configurational entropy can be calculated using a simple approximation known as the self-consistent phonon method. The method originally dates to Fixman \cite{fixman_highly_1969} and has since been applied to the hard sphere liquid \cite{stoessel_linear_1984}, a network liquid with hard sphere repulsions \cite{hall_microscopic_2003}, and a Lennard-Jones liquid \cite{hall_intermolecular_2008}. Very similar approximations are made in the replica theory of the hard sphere glass in high dimensions \cite{kurchan_exact_2012,kurchan_exact_2013,charbonneau_exact_2014}. The goal of self-consistent phonon theory is to compute separately the free energies of a frozen glass state configuration and the liquid state ensemble. The configurational entropy can be inferred from the difference in free energies. The core approximation used in self-consistent phonon theory is that each particle vibrates around its center of mass $\bm R_i$ and contributes an approximately Gaussian density profile $\rho_i(\bm r)=(\alpha_i/\pi)^{3/2}\exp(-\alpha_i(\bm r-\bm R_i)^2)$. The thermal vibration tensor $\bm{\alpha_i}=\frac{2}{3}\langle(\bm r_i-\bm R_i)^2\rangle^{-1}$ is determined from the local effective potential $\exp(\beta V_i^\mathrm{eff})=\prod_{j\neq i}\int d\bm{r_j} \rho_j(\bm{r_j}) \exp\left(-\frac{\beta}{2} V_{ij}(r_j-R_i)\right)$ where $V_{ij}$ is the interparticle interaction. At a structurally averaged mean-field level, the product reduces to an integral over the radial distribution function so that $
\alpha=\frac{\rho}{6}\int 4\pi R^2dR g(R) \nabla^2 V^\mathrm{eff}(R)$. Once $\alpha$ is known, the free energy per particle of the glassy state follows as 
\begin{eqnarray}
    f_g=&k_B T\rho\int d\bm{R} g(R;\rho) \sum_n p_n V^\mathrm{eff}(\bm{R};\alpha_n)\nonumber\\
    &-k_BT \sum_n p_n \frac{3}{2}\ln \frac{\pi}{\alpha_n}-Ts_\mathrm{mix}
\end{eqnarray}
where generally, there may be different types of particles with different $\alpha$ values, indexed by $n$. In our case, particles with $n$ bonds have a vibration tensor $\alpha_n$. $s_\mathrm{mix}$ is the mixing entropy for particles of different bond number. A similar expression has also been proposed by Sciortino et al. \cite{sciortino_inherent_1999} to describe the liquid in terms of the free energies of ``inherent structures'' of the liquid. Note that in our present model, the mixing entropy may be accessed without rearranging particles (i.e. in a single glassy state) due to rotational diffusion, which can cause bonds to break in some locations and form in others. 

The model used in the present work describes pairs of particles as interacting through a harmonic bonding potential $V_{ij}=-\epsilon_b +\frac{1}{2}\kappa_b (r-d_b)^2$ favoring a certain bond length $d_b$ and through a non-bonding Lennard-Jones interaction $V_{ij}=4\epsilon_{LJ}\left[\left(\frac{\sigma_{LJ}}{R}\right)^{12}-\left(\frac{\sigma_{LJ}}{R}\right)^{6}\right]$ capturing the steric core repulsion. Throughout this paper, we take units where $\sigma_{LJ}=1$. In this paper, we will also restrict our discussion to the case $d_b=d_{HS}$, although it is clear that the repulsions and bonding interactions have different origins and generally will have different length scales. When $d_b> \sigma_{LJ}$, or more generally when the microscopic interactions have multipli length scales, a variety of equilibrium phases can emerge even in the liquid state, leading to ``polyamorphism'' \cite{franzese_generic_2001}. We take $\kappa_b d_b^2/k_BT=300$, which is a value appropriate for silica potentials. This is the spring constant used by Hall and Wolynes \cite{hall_microscopic_2003}. Notice also the present model neglects both angular bonding constraints and the typically composite nature of the glass's components.

In this paper we go beyond the Hall-Wolynes model to consider a mixture of two type of particles: ``non-bonding particles'' with a mole fraction $x$ and ``bonding particles'' with mole fraction $1-x$. Non-bonding particles interact only via the non-bonding Lennard-Jones interactions while bonding particles may additionally form up to 4 bonds with other adjacent bonding particles. We also will allow the bonds to break thermally. Due to thermal fluctuations only some fraction $\Gamma$ of the maximum number of bonds will be formed on average at any given moment. Following Hall and Wolynes \cite{hall_microscopic_2003} and Erukhimovich \cite{erukhimovich_theory_2001}, we assume that bonds are formed independently, so that the probability that a given bonding particle has $n$ bonds is $p_n^b=\binom{4}{n}\Gamma^n(1-\Gamma)^{4-n}$. The overall probability that a particle has $n$ bonds is then 
\begin{equation}
    p_n(\Gamma,x)=
    \begin{cases}
        x+(1-x)p_n^b(\Gamma),& n=0\\
        (1-x)p_n^b(\Gamma), & n\neq 0
    \end{cases}
\end{equation}

To apply the self-consistent phonon method, we calculate first the radial distribution function $g(R;\rho)$. We assume that the structure of the liquid comes primarily from the repulsive interactions of the Lennard-Jones interactions. This assumption, of course, fails to account for the tetrahedral nature of substances such as silica. The radial distribution function is approximated by the Wertheim \cite{wertheim_exact_1963} solution to the Percus-Yevick equation, with a correction due to Verlet and Weis \cite{verlet_equilibrium_1972}. This radial distribution function $g_{HS}(R,\rho_{HS}^\mathrm{eff})$ depends on an effective hard sphere density $\rho_{HS}^\mathrm{eff}=\rho d_{HS}^3(T)$. One method for computing the effective hard sphere diameter $d_{HS}(T)$ is due to Barker and Henderson \cite{barker_perturbation_1967}, where the hard sphere diameter follows the equation 
\begin{equation}\label{eq: Barker-Henderson}
    d_{HS}=\int_0^{2^{1/6}} \left(1-\exp(-u(R)/k_BT)\right)dR.
\end{equation}
$u(R)$ is the repulsive part of the pair potential which reaches a value $u(2^{1/6})=0$. For Lennard-Jones particles, the repulsive part of the pair potential can be written
\begin{equation}
    u(R)=\begin{cases}
        4\epsilon_{LJ}\left[\left(\frac{1}{R}\right)^{12}-\left(\frac{1}{R}\right)^{6}\right]+\epsilon_{LJ} &R<2^{1/6}\\
        0 & R\geq 2^{1/6}
    \end{cases}
\end{equation} This rather picturesque choice of separating the Lennard-Jones potential into attractive and repulsive parts based on the force is due to Weeks, Chandler, and Andersen \cite{andersen_relationship_1971}, although other choices \cite{kang_perturbation_1986} have been found to yield similar results and to be superior at high density \cite{kang_perturbation_1986}. Throughout this paper, $\rho$ refers to the total number density of particles $\rho=N/V$ while $\rho_{HS}^\mathrm{eff}$ refers to the effective hard sphere density used to calculate $g(R)$.

Once we know the free energy per particle of a glassy state, we next need to compute the free energy per particle $f_l$ of the liquid state ensemble at the same density and bond distribution (i.e. $\Gamma$ and $x$). Once we know both free energies, we can determine the configurational entropy from the difference $f_l\approx f_g-Ts_c$. We use the following formula for the liquid state free energy per particle:
\begin{eqnarray}\label{eq: fliquid}
    f_l=&k_B T \ln(\rho/e)+k_B T \Psi\left(\rho_{HS}^\mathrm{eff} \right)+f_{scp}^b (\alpha\rightarrow 0)\nonumber\\
    &-\epsilon_b 2\Gamma (1-x)-Ts_\mathrm{mix} (x,\Gamma)-a\rho
\end{eqnarray}

The first term is the ideal gas entropy and the second is the interaction free energy due to the repulsive interactions. We assume the repulsive interactions can be represented by an effective hard sphere interaction, and use a formula due to Wang, Khoshkbarchi, and Vera \cite{wang_new_1996} for the interaction free energy of the hard sphere liquid. The formula expands the compressibility factor as a power series $Z(\rho_{HS})=\frac{1}{\xi}\sum_i D_i\left(\frac{\xi}{1-\xi}\right)^i$ where $\xi=\frac{1}{\sqrt{2}}\rho_{HS}$ and the coefficients $D_i=1, 1.96192, 0.55927, - 1.10721, 0.55626, -0.11923, 0.00954...$ are chosen to match the virial coefficients of the hard sphere liquid. Terms past $i=7$ are truncated. To find the contribution of repulsive forces to the liquid free energy, we simply integrate $\Psi(\rho_{HS})=\int_0^{\rho_{HS}} \frac{d\rho'}{\rho'} Z(\rho')$

The net entropy change due to bonding $f_{scp}^b(\alpha\rightarrow0)$ takes the form \cite{hall_microscopic_2003} $f_{scp}^b(\alpha\rightarrow0)=-k_B T(1-x)2\Gamma \ln\left[\left(\frac{2\pi}{\beta\kappa^b d_b^2}\right)^{1/2}\left(\rho/e\right)^{1/3}\right]$. Each bond formed reduces the energy by an amount $\epsilon_b$, thus the total energy due to bonding is $-\epsilon_b 2\Gamma (1-x)$. There is an additional entropy due to mixing of particles with different bond numbers, which takes the form $s_\mathrm{mix}/k_B=-x\ln x-(1-x)\ln(1-x)-2(1-x)[\Gamma\ln\Gamma+(1-\Gamma)\ln(1-\Gamma)]$. Since we are going to explore the liquid's properties both as a function of density and pressure, we will approximate the effect of the attractive nonbonding interactions using a Van der Waals term $-a\rho$. The last three terms are present in both $f_l$ and $f_g$, and thus cancel out when calculating $s_c$, but are important in the equation of state, thereby allowing us to compare constant volume and constant pressure experiments.

\subsection{Library construction for network glasses}

\begin{figure*}
    \centering
    \includegraphics[width=\textwidth]{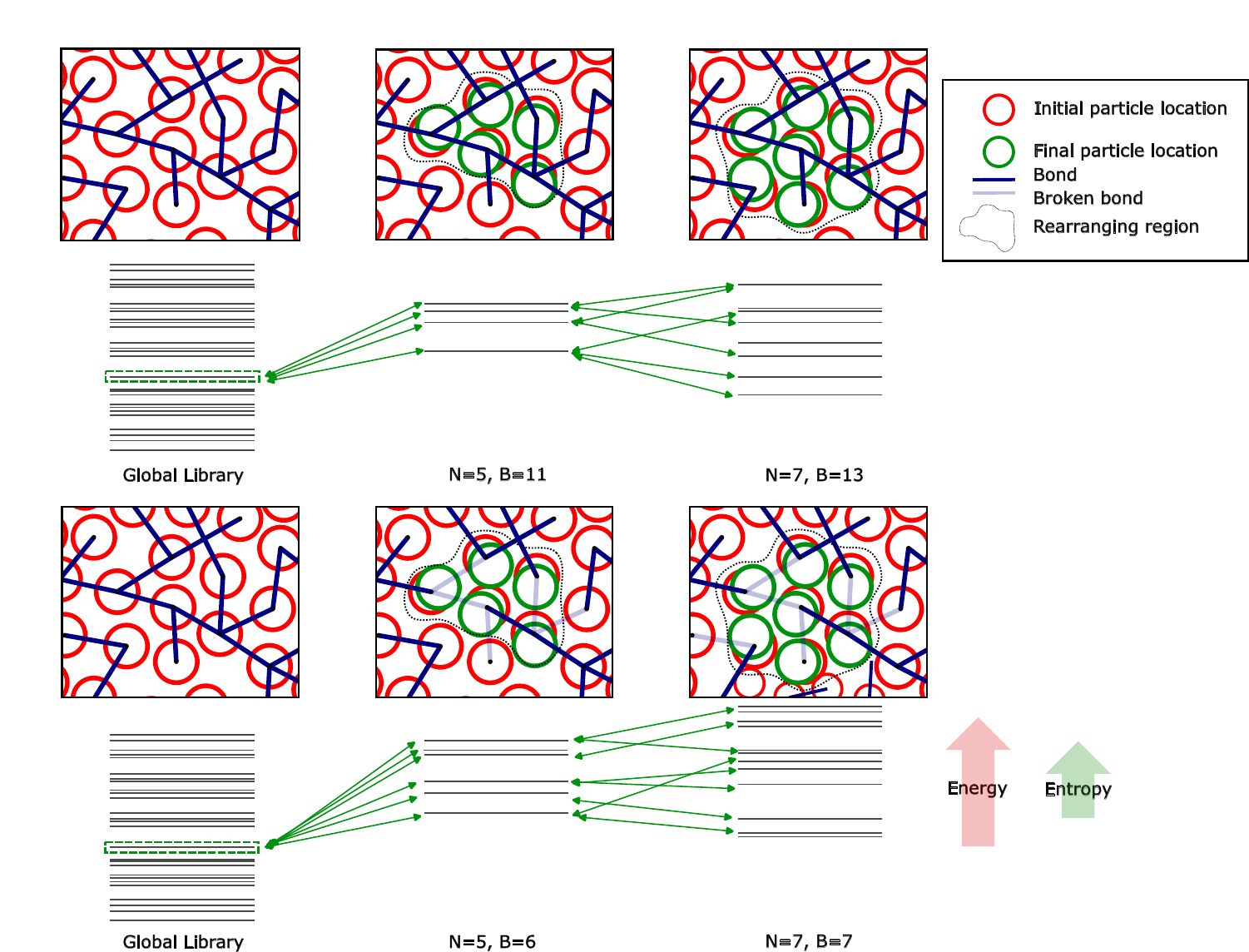}
    \caption{This schematic summarizes the library construction of the RFOT theory as it applies to network glasses. On the left hand side, the system may start in one of many possible initial states, each of which is a local minimum of the free energy landscape. These states are each part of the ``global library'', which tracks the free energy of each possible linearly stable state. To construct a ``local library'', we consider freezing all particles outside a local region with $N$ particles, but allowing them to elastically deform while allowing the particles in the local region to rearrange freely. The resulting ``rearranging droplet'' samples states which resemble a subset of the global library which makes up the local library. These states have higher energy on average than the initial state because the mismatch between the rearranged droplet region and the surrounding fixed glassy state near the boundary of the droplet. As the droplet grows, both the number of states accessed and the mismatch energies grow. When the droplet is sufficiently large, it will rearrange to a state with the same free energy as the initial state and thus can remain in the new state. At this droplet size, the configurational entropy balances with the mismatch energy. A rearrangement can occur without breaking any bonds (as shown in the top row) or while breaking some bonds (as shown in the bottom row). Breaking bonds requires an input of free energy, so the states will have higher energy on average than if no bonds were broken.}
    \label{fig:library}
\end{figure*}

In its modern formulation, the RFOT theory describes the dynamics of a supercooled liquid in terms of a construction of libraries of glassy states \cite{lubchenko_theory_2004, bouchaud_adam-gibbs-kirkpatrick-thirumalai-wolynes_2004}. The ``global library" includes each glassy configuration and records the energy and vibrational entropy of the state. If a system starts in a specific glassy configuration, and we “freeze” all particles outside a region of size $N$, then rearrangements of the droplet of $N$ particles will result in a set of locally stable structures known as a ``local library”, as shown in figure \ref{fig:library}. In a region with $N$ particles and at a temperature and density where the configurational entropy per particle is $s_c$, there are $\Omega=\exp(Ns_c/k_B)$ glassy configurations sampled. The local library is called local because the boundary of the droplet is confined in space, and thus all rearrangements are due to local motion. Similar protocols have been directly implemented in simulations \cite{cammarota_novel_2007} to determine the point-to-set correlations. Here we will use the library construction as a gedanken experiment rather than an explicit computer construction. Since rearrangements in a supercooled liquid occur locally, the local libraries catalog which rearrangements can occur. Once a rearrangement has occurred, the local libraries near the location of the rearrangement are updated to reflect the new local environment. Updating the local libraries in this way particularly has the effect of ``facilitating'' motion in regions with statistically higher energy barriers \cite{bhattacharyya_facilitation_2008}.

Once a local library is specified, it is used to construct a rearrangement free energy profile $F(N)$ that describes the thermodynamic cost of rearranging a region of size $N$. $N$ functions as a collective local rearrangement reaction coordinate, much as in ordinary nucleation theory. The rate of the rearrangement depends on progressing across the free energy barrier $\Delta F^\ddag$, the maximum value of $F(N)$. Suppose the $j$th glassy state has $n_b$ bonds per particle and a free energy $\phi_{j,n_b}^\mathrm{lib}$, which has components related to the energy from intermolecular forces (including bonding, attractive, and repulsive forces), the vibrational entropy of particles vibrating around their fixed points, and a mixing entropy-like term associated with the number of possible arrangements of the $N n_b/2$ bonds. When a state $\phi_{j,n_b}^\mathrm{lib}$ is locally inserted into a global frozen structure, the free energy cost will be higher than the difference in bulk free energies by an amount $\Gamma_{j,in}$ because the two structures do not match at their boundaries. That is, 

\begin{equation}\label{eq:lib diff}
    \phi_{j,n_b}^\mathrm{lib}-\phi_{in}^\mathrm{lib}=\Phi_{j,n_b}^\mathrm{bulk}-\Phi_{in}^\mathrm{bulk}+\Gamma_{j,in}
\end{equation}

The ``mismatch energy'' accounts for the fact that the initial state already matched the boundary conditions of the rearranging region more favorably than other still typical states in the local library, while the new configuration will not. The na{\"i}ve scaling $\Gamma_{j,in}=\gamma N^{2/3}$ found at ordinary first order transitions is not accurate near the ideal glass transition due to the multiplicity of states accessible to the liquid. These other states allow a more gradual transition in interface structure. Dzero, Schmalian, and Wolynes \cite{dzero_activated_2005} have shown that in the replica treatment of these droplets, high order replica symmetry breaking is expected in the interface. Near the interface of the rearranging region, if the states inside and outside the droplet are initially poorly matched, then a third configuration from the library can be interpolated between them to reduce the energy cost much as wetting occurs in multiphase liquid-gas transitions. Using an argument based on the ideas of Villain \cite{villain_equilibrium_1985} for the random field Ising model suggests that the scaling should be instead\cite{kirkpatrick_scaling_1989} $\Gamma_{j,in}=\gamma N^{1/2}$  near $T_K$ in a random first order transition. As Kirkpatrick, Thirumalai, and Wolynes show, this scaling in turn would ensure there is only one length scale that grows near an ideal glass transition.

Density functional theory provides an estimate of the mismatch energy in terms of the entropic cost of localizing a particle near the interface \cite{xia_fragilities_2000}. Using this estimate, Xia and Wolynes inferred that near $T_K$, $\gamma=\gamma_0 k_B T=\frac{3\sqrt{3\pi}}{2}k_BT\ln\left(\frac{(a/d_L)^2}{\pi e}\right)$. In density functional calculations the Lindemann ratio $d_L/a\approx 0.1$ takes on a nearly universal value of 0.1. This value is very close to what would be inferred from the plateau seen in neutron scattering. Since the mismatch parameter $\gamma$ depends only logarithmically on the Lindemann ratio, the approximation that $\gamma_0=$constant would appear to be quite good for many substances. In principle, of course, the plateau and the mismatch energy itself could depend on details of the bonding potentials since they introduce new microscopic length scales.

Near the onset temperature for activated dynamics $T_A>T_K$, the mismatch energy vanishes because $T_A$ is a spinodal. Dzero, Schmalian, and Wolynes discuss this softening in replica theory \cite{dzero_activated_2005}. This ``softening'' effect can be incorporated in the molecular case using a correction to the mismatch energy \cite{lubchenko_barrier_2003}. Near $T_A$, the mismatch energy is approximately $\Sigma_A=\gamma_{LW}N^{2/3}$, where $\gamma_{LW}(T)$ is a coefficient which may be calculated from a Landau-Ginzburg functional with a spinodal transition. $\gamma_{LW}(T)$ vanishes when $T=T_A$. On the other hand, the mismatch near $T_K$ is $\Sigma_K=\gamma_0 k_B T N^{1/2}$, as discussed earlier. The mismatch in the $T\rightarrow T_K$ and $T\rightarrow T_A$ limits can be combined in a simple ansatz, where $\Sigma=\frac{\Sigma_A\Sigma_K}{\Sigma_A+\Sigma_K}$.

Lubchenko and Wolynes \cite{lubchenko_barrier_2003} combined the two limits to give the global dependence of the barrier on temperature. $T_A$ was treated by them as a fitting parameter when considering softening effects in experimental viscosity data. Later, Stevenson, Schmalian, and Wolynes \cite{stevenson_shapes_2006} showed that $T_A$ could itself be estimated since local rearrangements in an equilibrium liquid change character  from being compact to stringy dramatically at a temperature $T_C$ which marks the crossover between regimes of qualitatively different rearrangement dynamics. The Stevenson, Schmalian, and Wolynes theory can be thought of as a microscopic way of inferring the spinodal temperature that goes beyond mean field theory. Above $T_C$, rearrangements are largely collisional in nature and well described by mode coupling theory. Below $T_C$, particles rearrange in local, collective activated events as described by the RFOT theory. Near $T_K$ the rearrangements are relatively compact but as $T_C$ is approached from below, rearrangements become more ramified and string-like in shape \cite{stevenson_shapes_2006}, until at $T_C$, the barrier to a string-shaped rearrangement vanishes. Notice this picture brings in a second length scale (the interface width) but both lengths are comparable in ordinary liquids near the laboratory $T_g$. The configurational entropy at the crossover was estimated by Stevenson, Schmalian, and Wolynes based on the shape entropy of strings or percolating ``lattice animals''. The string transition occurs at a nearly universal value of the configurational entropy $s_c^\mathrm{string}=1.13 k_B$ or $s_c^\mathrm{perc}=1.28 k_B$ depending on the method. These transition configurational entropies per particle are universal as long as they are counted per ``bead'' \cite{stevenson_thermodynamickinetic_2005}, based on the effective spherical rearranging units in the particular liquid. In this paper, rather than using $T_A$ as a fitting parameter as Lubchenko and Wolynes did, we will approximate $T_A\approx T_C$ using the percolation cluster estimate.

At this point, we see that all the microscopic details are available to allow us to use the free energy of the library states to calculate the free energy for rearranging a region of size $N$ from an initially equilibrated glassy state for models of bonded liquids. Since the initial state can rearrange to any of the $\Omega=\exp(Ns_c/k_B)$ other states in the local library, the rearrangement free energy is $F(N,B)=-T s_c(B/N)N+\phi_{j,n_b}^\mathrm{lib}-\phi_{in}^\mathrm{lib}$ where $B=n_bN/2$ is the number of bonds in the rearranging droplet and we have written explicitly the dependence of $s_c$ on the number of bonds per particle. As mentioned in equation \ref{eq:lib diff}, the free energy difference in individual library states $\phi_{j,n_b}^\mathrm{lib}-\phi_{in}^\mathrm{lib}$ includes both bulk and mismatch terms. The bulk free energy of a glassy state can be calculated using the self-consistent phonon method. It is simplest to combine thoese contributions with the configurational entropy when writing the free energy of a local rearrangement event.
\begin{subequations}
\begin{eqnarray}\label{eq:FNB}
    F(N,B)&=&-TNs_c(B/N)+\Sigma(N,s_c)\nonumber\\
    &&+Nf_g(B/N)-Nf_g(B_{in}/N)\\
    F(N,B)&=&-TNs_c(B_{in}/N)+\Sigma(N,s_c)\nonumber\\
    &&+Nf_l^b(B/N)-Nf_l^b(B_{in}/N)
\end{eqnarray}
\end{subequations}

where we have averaged over library states with different numbers of bonds to replace $\Phi_x^\mathrm{bulk}$ with $f_x(B/N)$, the mean free energy per particle in a glass or liquid state with $B/N$ bonds per particle. Note that a liquid held at a given temperature and pressure will undergo fluctuations in the number of bonds $B$ according to the free energy $\delta F^b=Nf_l^b(B/N)-Nf_l^b(B_{in}/N)$. The mean fraction of bonds formed at any given time (i.e. $\Gamma$) is chosen to minimize $\delta F^b$ according to $B=2\Gamma (1-x) N$. Explicitly, we can write 

\begin{subequations}
\begin{eqnarray}
    f_l^b(\Gamma)&=&\left(-\epsilon_b -Ts_b(\rho)\right)2\Gamma (1-x)-Ts_\mathrm{mix}(x,\Gamma)\\
    s_b/k_B&=&\log\left[\left(\frac{2\pi}{\beta \kappa^b d_b^2}\right)^{1/2}\left(\frac{\rho}{e}\right)^{1/3}\right]\\
    s_\mathrm{mix}/k_B&=&-x\ln x-(1-x)\ln(1-x)\nonumber\\
    &&-2(1-x)\left[\Gamma \ln\Gamma+(1-\Gamma)\ln(1-\Gamma)\right]
\end{eqnarray}
\end{subequations}

which is minimized by 
\begin{equation}
    \Gamma=\frac{1}{1-\exp\left(-\epsilon_b/k_BT-s_b\right)}
\end{equation}

\begin{figure*}
    \centering
    \includegraphics[width=\linewidth]{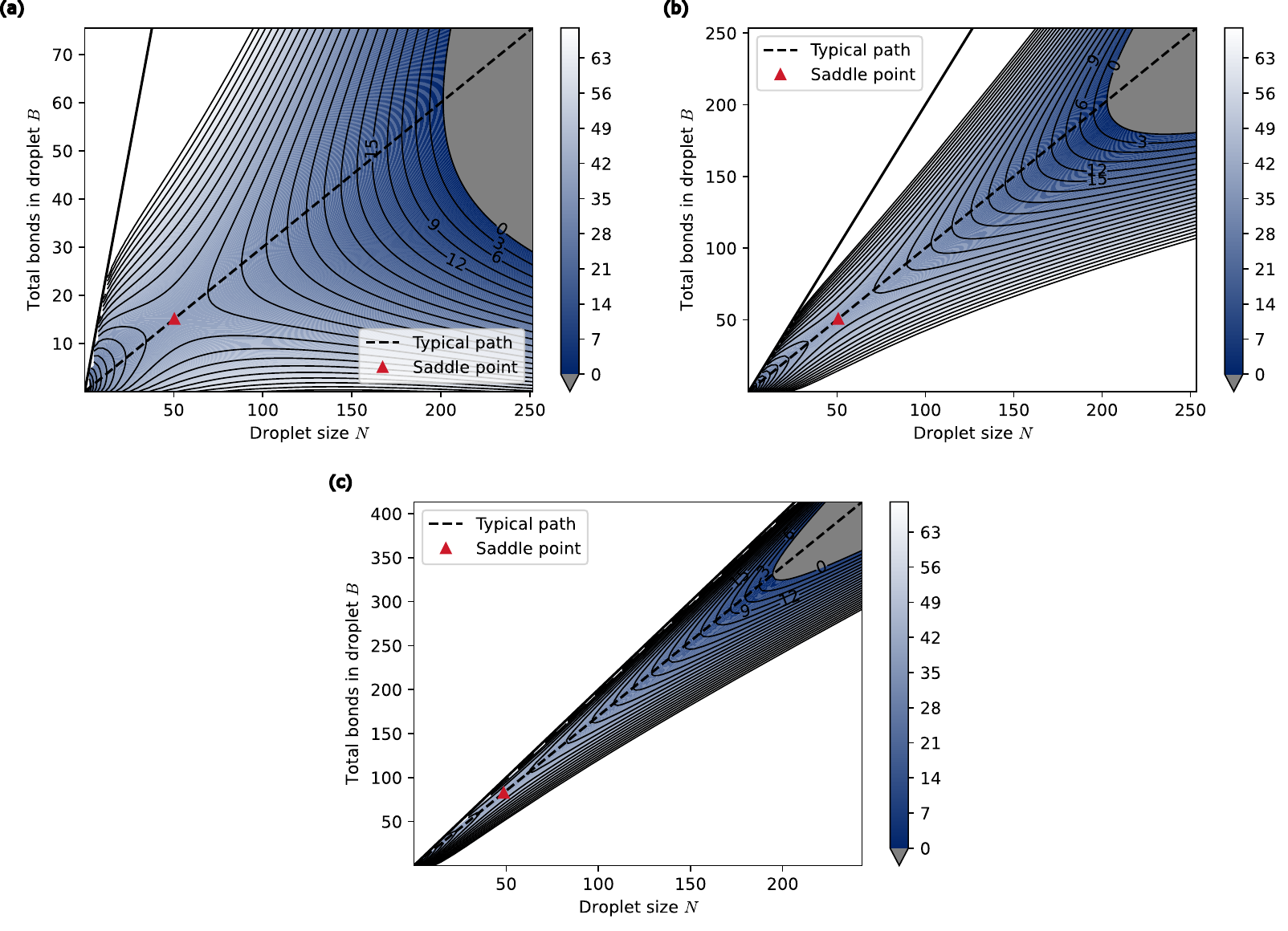}

    \caption{Rearrangement free energy as a function of the size of a rearranging droplet, $N$, and the number of bonds in the droplet, $B$. Each plot is close to $T_g$, and has no non-bonding impurities $(x=0)$. Bond energies $\epsilon_b$ are chosen so that (a) 15\%, 50\%, and 85\% of the maximum possible number of bonds are formed at equilibrium in the three cases. We have included no non-bonding impurities $(x=0)$. In each plot, the free energy in units of $k_BT$ is shown by the shade of blue with level curves every $3k_BT$. The grey region in the upper right indicates where $F<0$. According to the RFOT theory, once a droplet has grown large enough to reach this region, the rearrangement is complete and the particles will locally freeze into a new glassy state. We see that the saddle point lies along the line where the number of bonds is equilibrated.}
    \label{fig:FNB}
\end{figure*}

Since we are considering rearrangements in an equilibrium liquid, $B_{in}$ will always be chosen to minimize $\delta F^b$, and thus $F(N,B)\geq F(N,B_{in})$. For aging network liquids with an out-of-equilibrium number of bonds, $\delta F^b$ will act as a contribution to the driving force for rearrangements. We plan to explore such glassy states with possibly nonequilibrium conditions of bonding chemical equilibrium elsewhere.

The rearrangement scenario considered by both Mott and Turnbull \cite{mott_viscosity_1987, de_neufville_effect_1970} has a simple interpretation in terms of our rearrangement free energy $F(N,B)$, which is plotted in figure \ref{fig:FNB} for three different $\epsilon_b$ values and temperatures near $T_g$. The rearrangement pictured by Mott and Turnbull corresponds to a particular path through the space of $N$ and $B$, where first $B$ increases through bond breaking, then $N$ increases as particles reconfigure, and finally $B$ decreases back to an equilibrium value by bonds reforming after the reconfiguration. This path does lead to a rearrangement, as it terminates in the gray region of the plots in figure \ref{fig:FNB} where $F(N,B)<0$, but it does so with a strictly higher free energy barrier than a simple path without any bond breaking.

Equation \ref{eq:FNB} is the analogue of $F(N)$ in the RFOT theory without bond breaking effects, and it should be interpreted in the same way. When a system evolves from one glassy state to another, it does so by first forming a liquid droplet of $N$ particles and $B$ bonds. $F(N,B)$ is the free energy cost of forming that droplet. The rate for glassy rearrangements can be determined from the free energy barrier calculated using $F(N,B)$.

\section{Connection with experimental measurements}
There are no inorganic materials that precisely coincide with the simple model presented here. Silica, for instance, has multiple atoms, angular constraints, and radial forces competing to influence the molecular dynamics. Nevertheless, we will make comparisons to two classes of inorganic network liquids in order to make sense of some trends in our model. We compare predictions of our model with data for alkali silicates \cite{bohmer_nonexponential_1993, poole_low-temperature_1949} and alkali germanates \cite{shelby_viscosity_1974}.

In experiments, the convenient theoretical quantities $\rho_{HS}^\mathrm{eff}$, $x$, and $\Gamma$ need to be related to laboratory constraints. Typically, the microscopic properties of the bonds in a given material determine a binding energy $\epsilon_b$ from which the bonded fraction $\Gamma$ is determined as a function of temperature. Experiments are most often performed at constant pressure while varying the temperature, $T$. Occasionally, high pressure techniques make the properties at fixed volume available for study. To compare predicted activation barriers as a function of $T$ to experiments performed at constant pressure, we determine the density naturally from an equation of state $\rho(P,\Gamma, T)$. For experiments performed at constant volume, the effective density of hard sphere-like units $\rho_{HS}^\mathrm{eff}$ of the liquid varies with temperature due to variations of the effective hard sphere diameter $d_{HS}(T)$, giving the main $T$ dependence

The temperature dependent effective hard sphere density can be used together with the calculated values for $s_c(\rho_{HS}^\mathrm{eff},\Gamma,x)$, which yields predictions for the viscosity $\eta$ and fragility $m=\left(\frac{\partial (\log \eta)}{\partial (T_g/T)}\right)_{T=T_g}$ \cite{angell_formation_1995} at constant volume or pressure and varying temperature.

\subsection{Constant volume protocols}

\begin{figure}
    \centering
    \includegraphics[width=\linewidth]{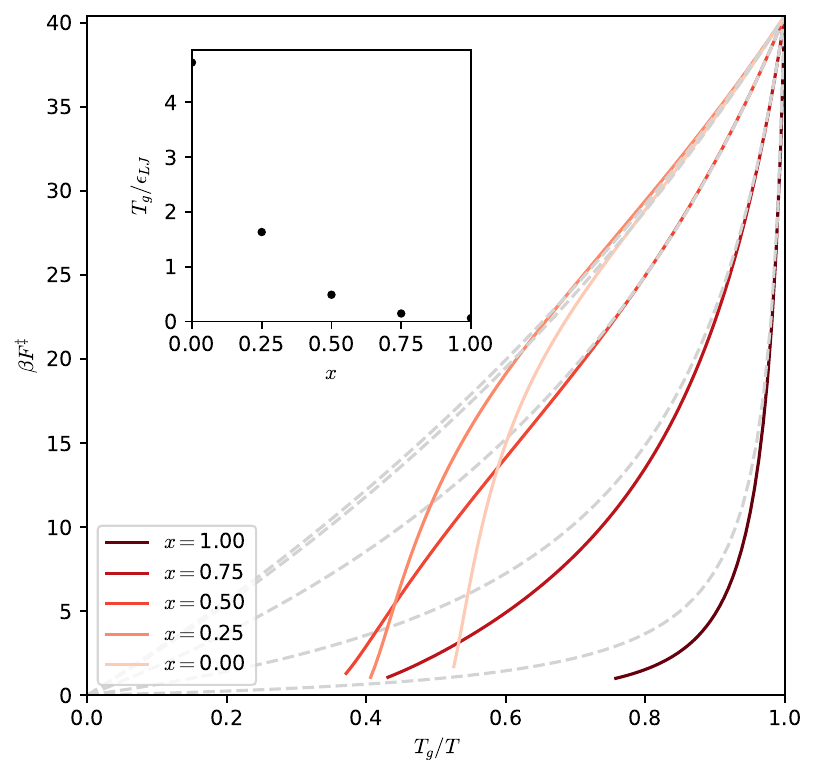}

    \caption{Barrier height for rearrangements vs $T_g/T$. All curves are shown for constant density $\rho_{LJ}=0.92$ and $\epsilon_b=50\epsilon_{LJ}$. This represents a liquid with very strong bonds and different compositions. Because $\epsilon_b$ is so high, nearly every bonding particle has 4 bonds. Here, the number density $\rho_{LJ}$ is chosen so that the completely nonbonded ($x=1$) liquid has a pressure $P=0$ at $T_g$. As we increase the fraction of particles which are non-bonding, the material becomes more fragile. Additionally, $T_g$ decreases with the amount of non-bonding impurities. For this plot, the barrier height was calculated by first finding the configurational entropy $s_c$ at a specific temperature and then maximizing equation \ref{eq:FNB}.We have plotted in dashed lines the VFT curves inferred based on the fragility parameter $m$ by matching the value and slope of the barrier height near $T_g$. At higher temperatures, near $T_C$, the barrier height separates from VFT-like behavior seen at lower temperatures and drops to zero due to softening effects.}
    \label{fig:Angell V,n}
\end{figure}

\begin{figure}
    \centering
    \includegraphics[width=\linewidth]{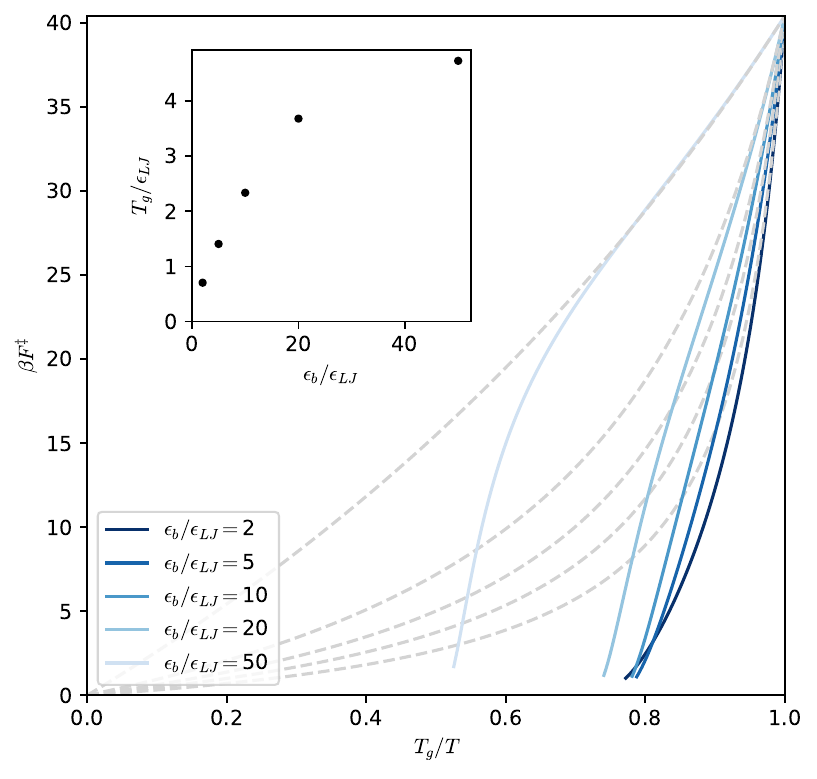}

    \caption{Barrier height for rearrangements vs $T_g/T$. All curves are shown for constant density $\rho_{LJ}=0.92$ and $x=0$. This represents a liquid of purely bonding particles and different bond strengths. Here, the number density $\rho_{LJ}$ is chosen so that the completely nonbonded ($x=1$) material has a pressure $P=0$ at $T_g$. As we increase the the strength of the bonds, the material becomes stronger. Additionally, $T_g$ increases with the strength of the bonds. For this plot, the barrier height was calculated by first finding the configurational entropy $s_c$ at a specific temperature and then maximizing equation \ref{eq:FNB}. We have plotted in dashed lines the VFT curves inferred based on the fragility parameter $m$ by matching the value and slope of the barrier height near $T_g$. At higher temperatures, near $T_C$, the barrier height separates from VFT-like behavior seen at lower temperatures and drops to zero due to softening effects.}
    \label{fig:Angell V,eps}
\end{figure}

In figure \ref{fig:Angell V,n}, we show plots of the static barrier height $\beta F^\ddag$ vs $T_g/T$. Plots of this form are often obtained experimentally by measuring the viscosity of a liquid at various temperatures and plotting log viscosity, assuming therefore a constant prefactor in the rates. We note, however, that the viscosity is only determined by the static barrier height below the crossover temperature $T_C$. At higher temperatures, mode coupling effects result in finite, temperature dependent viscosity despite the vanishing static barrier \cite{bhattacharyya_facilitation_2008,stevenson_shapes_2006}. An Arrhenius process has a rate which appears linear when plotted in this fashion, and thus this type of plot is often called an Arrhenius plot. The slope of the log viscosity curve near $T_g$ is the fragility $m$ \cite{angell_formation_1995}. The viscosity of strong liquids obeys a near-Arrhenius temperature dependence, and thus appears linear on an Arrhenius plot. By definition, all liquids have the same viscosity near $T_g$ and it appears that all liquids have a nearly universal viscosity at high temperatures given by collision results \cite{maxwell_illustrations_1860}. As a result, strong liquids have an $m$ value of around 17. In contrast, the viscosity of a fragile liquid instead shows visible curvature and a much higher slope near $T_g$ than strong liquids. VFT curves are shown in figure \ref{fig:Angell V,n} with dashed lines as guides for the eye. In all cases, we see deviations from a VFT fit at high temperatures due to interface softening.

Crudely speaking, tuning $x$ in the laboratory amounts to tuning the composition of silica through the addition of non-bonding impurities, such as happens in glasses formed by the common alkali silicates. In this model, the properties of the bonds (namely, $\epsilon_b$) stay fixed while the mole fraction of impurity $x$ is tuned. The trends of figure \ref{fig:Angell V,n} are similar to what is seen in real systems. $T_g$ increases as the number of bonds is increased, since the bonds introduce constraints, reducing the configurational entropy without cooling. Additionally, highly bonded systems tend to be stronger than non-bonded systems, and indeed here our model shows a transition from strong to fragile as $x$ is increased from 0 to 1.

In addition to tuning $x$ in the model, we can also tune the strength of bonds, $\epsilon_b$. Such bond tuning can not be reasonably implemented for molecular liquids, since it would require changing the microscopic properties of a bond. One could plausibly tune $\epsilon_b$ in patchy colloid experiments or in simulations, which would allow a direct test of our predictions. The results of tuning $\epsilon_b$ while keeping $x=0$ fixed are shown in figure \ref{fig:Angell V,eps}. $T_g$ increases with increasing $\epsilon_b$, reflecting the configurational entropy reduction caused by an increase in bond number per particle. Additionally, the fragility decreases as $\epsilon_b$ increases.

\subsection{Constant pressure protocols}

\begin{figure}
    \centering
    \includegraphics[width=\linewidth]{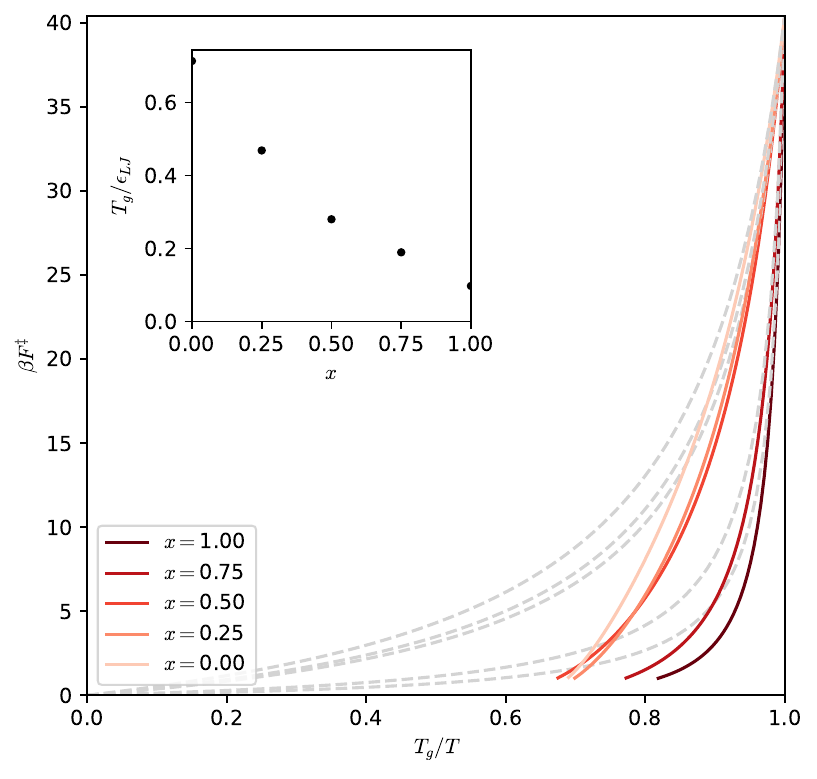}

    \caption{
    Barrier height for rearrangements vs $T_g/T$. All curves are shown for constant pressure $P=0$ and $\epsilon_b=50\epsilon_{LJ}$. This represents a liquid with very strong bonds and different compositions. Because $\epsilon_b$ is so high, the bonds are nearly completely saturated. As we increase the fraction of particles which are non-bonding, the material becomes more fragile. Additionally, $T_g$ decreases with the amount of non-bonding impurities. For this plot, the barrier height was calculated by first finding the configurational entropy $s_c$ at a specific temperature and then maximizing equation \ref{eq:FNB}. We have plotted in dashed lines the VFT curves inferred based on the fragility parameter $m$ by matching the value and slope of the barrier height near $T_g$. At higher temperatures, near $T_C$, the barrier height separates from VFT-like behavior seen at lower temperatures and drops to zero due to softening effects.}
    \label{fig:Angell P,n}
\end{figure}

\begin{figure}
    \centering
    \includegraphics[width=\linewidth]{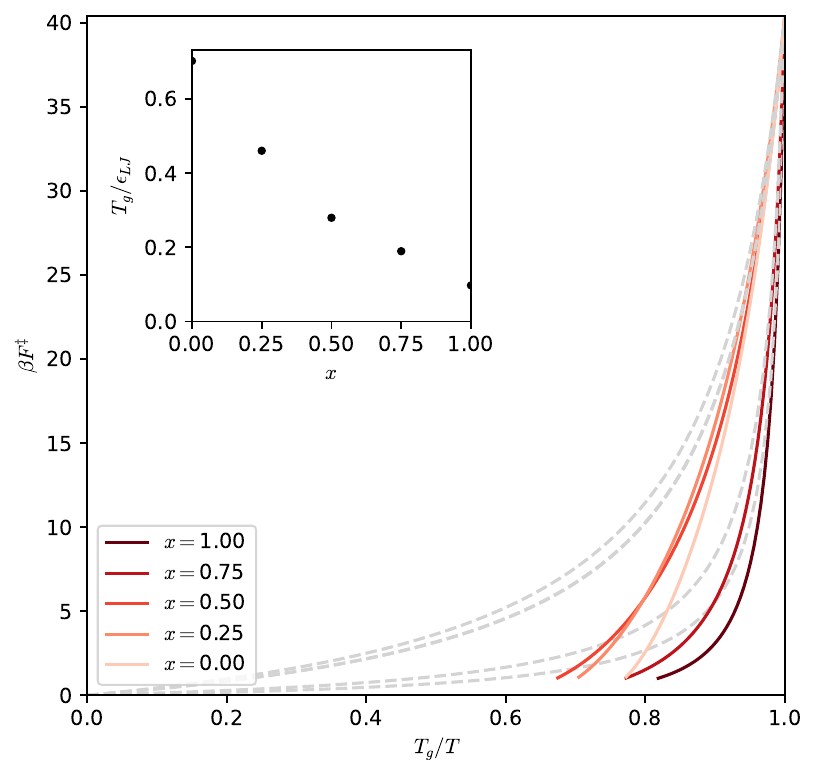}

    \caption{
    Barrier height for rearrangements vs $T_g/T$. All curves are shown for constant pressure $P=0$ and $\epsilon_b=5$. This represents a liquid with only moderately strong bonds and different compositions. Because $\epsilon_b$ is lower, the number of bonds is changing with temperature. As we increase the fraction of particles which are non-bonding, the liquid briefly becomes more strong up to a maximum at around $x=0.25$ before becoming more fragile with the addition of even more non-bonding particles. Additionally, $T_g$ decreases with the amount of non-bonding impurities. For this plot, the barrier height was calculated by first finding the configurational entropy $s_c$ at a specific temperature and then maximizing equation \ref{eq:FNB}. We have plotted in dashed lines the VFT curves inferred based on the fragility parameter $m$ by matching the value and slope of the barrier height near $T_g$. At higher temperatures, near $T_C$, the barrier height separates from VFT-like behavior seen at lower temperatures and drops to zero due to softening effects.}
    \label{fig:Angell P,n 5}
\end{figure}
To describe constant pressure experiments, we first construct the following equation of state:
\begin{equation}\label{eq: eos}
    \frac{P}{k_BT}=\rho\left[1-(1-x)\frac{2\Gamma}{3}\right]+\rho\left[Z\left(\rho_{HS}^\mathrm{eff}\right)-1\right]-\frac{a}{k_BT}\rho^2
\end{equation}
which can be obtained by differentiating equation \ref{eq: fliquid} with respect to volume. $Z$ is the compressibility factor \cite{wang_new_1996} associated with the repulsive nonbonding interactions. We note that some real network liquids exhibit both a ``high-density, unbonded" liquid state (HDL) and a ``low-density, bonded" liquid state (LDL) \cite{angell_liquid_2002, saika-voivod_free_2004}. There is no liquid-liquid phase transition in the present model, but such a transition could be modeled with an additional explicit bond dependence of the effective particle diameters in the equation of state. Within the present model, even the fully bonded liquid ($x=0,\Gamma=1$) has around $11$ nearest neighbors in the first peak of $g(R)$ while silica is notably still rather tetrahedral. For this reason, the present model has a density comparable to what one would expect for the high density amorphous phase of $\mathrm{SiO_2}$, rather than its more usual low density open structure. At a given pressure, we use equations \ref{eq: Barker-Henderson} and \ref{eq: eos} to compute $\rho_{HS}^\mathrm{eff}(T)$ and in turn $s_c(T)$. 

In figure \ref{fig:Angell P,n}, we plot the viscosity, glass transition temperature, and fragility for the present model at constant pressure. In this case, the liquid is held at constant pressure for several values of $x$. The plots here resemble those at constant volume in figure \ref{fig:Angell V,n}. Note that the fragility measured at constant pressure is significantly larger than the fragility measured at constant volume. This is true for any value of $x$. In real materials, typically $m_p>m_v$ as well, since 
\begin{gather}
    m_p-m_v=\left(\frac{\partial(\log \eta)}{\partial \rho} \right)_T \left(\frac{\partial\rho}{\partial T} \right)_P
\end{gather}
The second factor is positive, as long as the liquid has a positive thermal expansion coefficient. This is true for most materials, with a notable exception of water and silica under some conditions \cite{white_thermal_1973}. In water, the negative thermal expansion at low temperatures is a signature of the underlying liquid-liquid phase transition \cite{lucas_fragile--strong_2019}. Both the liquid-liquid phase transition and the negative thermal expansion could be incorporated into the model by means of a more accurate equation of state with explicit bond dependence. Real silica has a remarkably low thermal expansion coefficient, and thus very little difference between $m_P$ and $m_V$. This is another weakness of applying our model to describe the alkali silicates without a more accurate equation of state.

Figure \ref{fig:Angell P,n 5} is similar to figure \ref{fig:Angell P,n} but with a lower assumed bond energy. For this lower value of the bond energy, the fragility depends non-monotonically on $x$, while as before, the glass transition temperature still decreases as non-bonding particles are added.

\subsection{Fragility of liquids with different bond energies and numbers}

\begin{figure*}
    \centering
    \includegraphics[width=\linewidth]{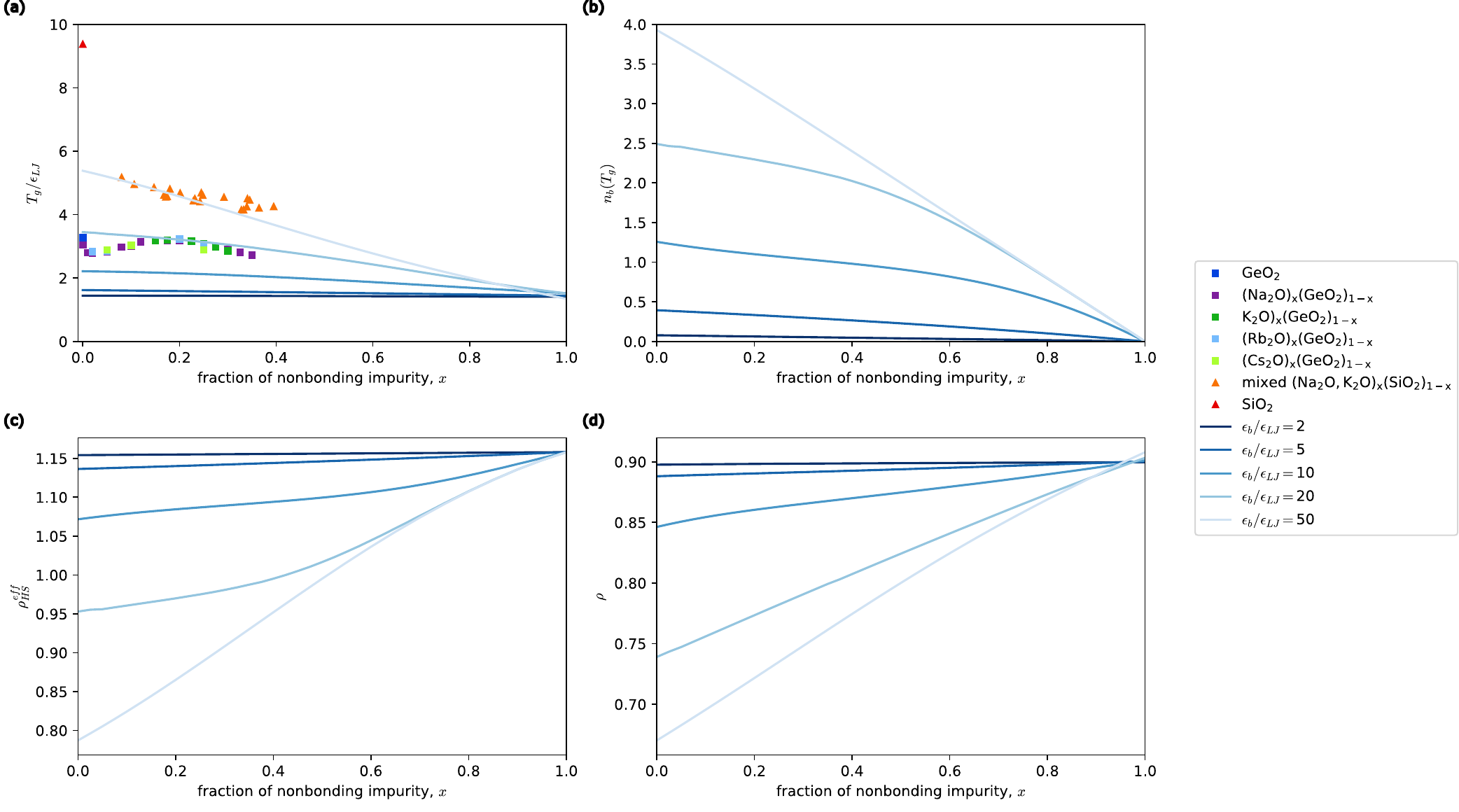}

    \caption{(a) Glass transition temperature, (b) bond number per particle, (c) effective hard sphere density and (d) number density at zero pressure and at $T_g$. Each curve is a different bond strength $\epsilon_b$ labeled by color, plotted as a function of the fraction of nonbonding impurity $x$. As we introduce more and more nonbonding impurities, $T_g$ decreases, since bonds reduce the configurational entropy. Naturally, as nonbonding impurities are introduced, the number of bonds per particle decreases. This trend is diminished for low $x$ and low $\epsilon_b$, since thermal bond breaking becomes significant when $\epsilon_b/T_g\sim1$. When $n_b(T_g)$ decreases, $\rho_{HS}^\mathrm{eff}$ must increase to offset the increase in configurational entropy. For any measurement, as $x$ gets close to 1, the bond strength becomes irrelevant since fewer and fewer particles are capable of forming bonds. Note therefore that each curve collapses onto one another near $x=1$. Points in (a) mark data for silicates \cite{bohmer_nonexponential_1993, poole_low-temperature_1949} and germanates \cite{shelby_viscosity_1974}. To compare the data to the model, we must choose an appropriate conversion factor between Lennard Jones units and Kelvins. For silicates, we choose $\epsilon_{LJ}=160$ K, and for germanates we choose $\epsilon_{LJ}=250$ K. The value of $\epsilon_{LJ}$ for silicates was chosen so that the data approximately fall on the $\epsilon_b/\epsilon_{LJ}=50$ curve. $T_g$ vs $x$ for silicates matches the high $\epsilon_b$ predictions of the model quite well, with the exception of the reported value for the $T_g$ of pure $\mathrm{SiO_2}$. Nearly pure silica has an exceptionally broad glass transition range that has been reported as being very sensitive to impurities \cite{bruning_characterization_2005}. Such a sensitivity does not appear in the present model and is quite puzzling to us. We note that the breadth of the transition may make the value hard to pin down. On the other hand, the present model does not show a sign of the weak nonmonotonicity of $T_g$ seen for the germanates.}
    \label{fig:Tg,Gamvsx}
\end{figure*}

\begin{figure*}
    \centering
    \includegraphics[width=\linewidth]{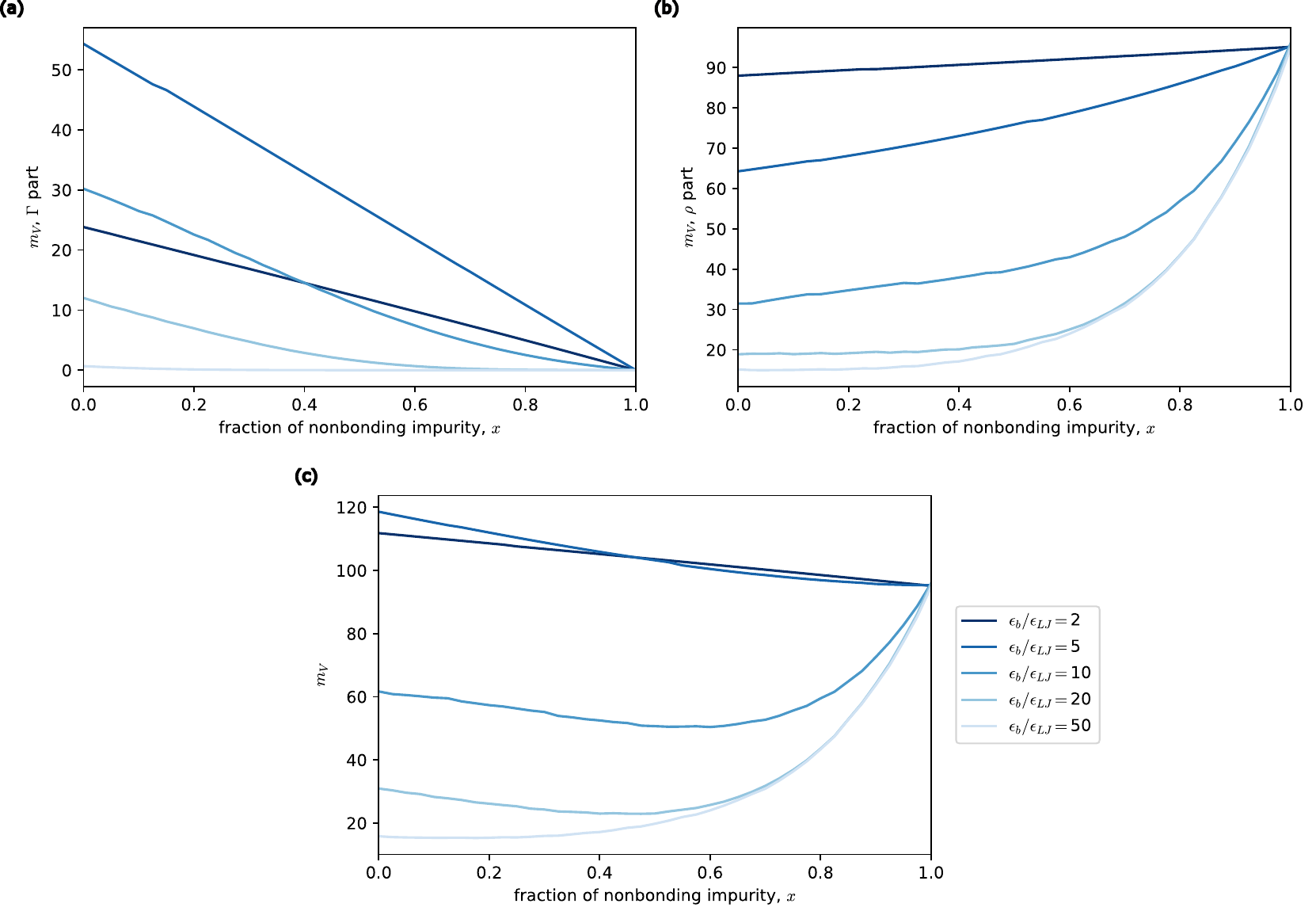}

    \caption{Here we show the fragility at constant volume, $m_V$. $m_V$ has components that arise from (a) changes in the number of bonds per particle $n_b$ and from (b) changes in the effective hard sphere density $\rho_{HS}^\mathrm{eff}$ with temperature. $m_V^{(n_b)}$ is negligible for strong bonds, as is likely true in silicates, but becomes relevant for the weaker bonds shown here. $m_V^{(\rho)}$ decreases with increasing $\epsilon_b$. (c) Adding the two contributions together yields the total fragility at constant volume. For sufficiently strong bonds $(\epsilon_b\gtrsim 20\epsilon_{LJ})$, the fragility increases monotonically with the amount of nonbonding impurity. For materials with weaker bonds $(\epsilon_b\lesssim 20\epsilon_{LJ})$, however,  the fragility is nonmonotonic with respect to bond number due to the competition between the two contributions to the fragility.}
    \label{fig:mvvsx}
\end{figure*}

\begin{figure*}
    \centering
    \includegraphics[width=\linewidth]{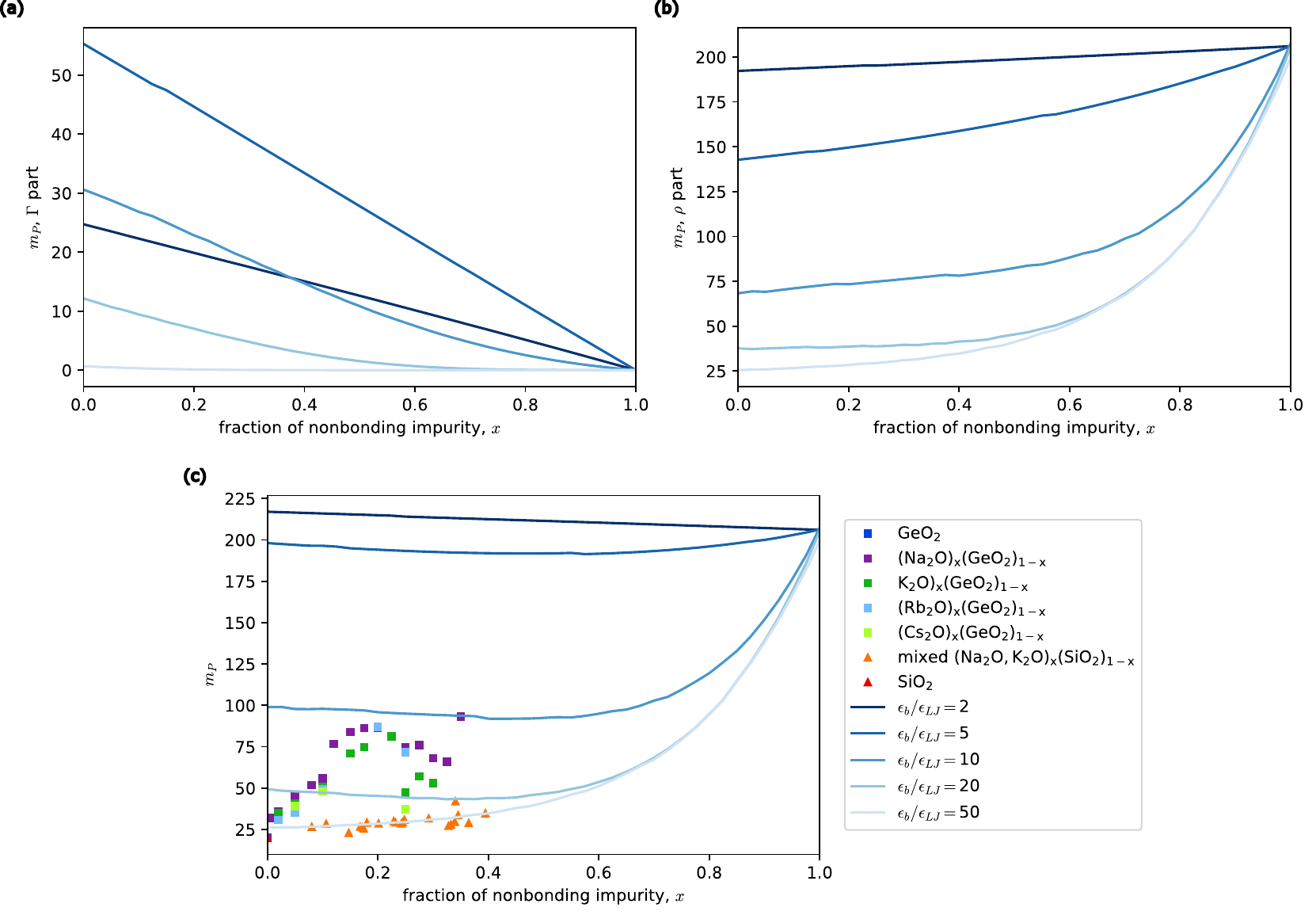}

    \caption{The fragility at constant pressure $m_P$ obeys many of the same trends as the fragility at constant volume $m_v$ shown in figure \ref{fig:mvvsx}. As before, the fragility can be split into (a) $m_P^{(n_b)}$ and (b) $m_P^{(\rho)}$. $m_P^{(n_b)}$ is roughly identical to $m_V^{(n_b)}$. $m_P^{(\rho)}$ is substantially larger than $m_V^{(\rho)}$, as is commonly found in glass forming liquids. As before, we have plotted data for silicates and germanates in (c). Our model accurately predicts the fragility of silicates across the measured range of compositions. It appears that the strength of the Si-O-Si bond in amorphous silicates is at least 50 times stronger than the effective non-bonding attractive forces. The model does not describe the nonmonotonicity of the germanates' fragility.}
    \label{fig:mpvsx}
\end{figure*}

In the previous section, we showed how to calculate the viscosity as a function of temperature, given a specific $x$, $\epsilon_b$, and $P$ or $\rho$. We saw that $T_g$ always increased with increasing numbers of bonds. The fragility, $m$ also depended on the number of bonds. For sufficiently strong bonds $(\epsilon_b/k_B T_g\gg 1)$, the fragility monotonically decreased with the number of bonds. This is quite consistent with the trend known experimentally in silica when non-bonding impurities, such as alkali oxides are added \cite{poole_low-temperature_1949}. Additionally, for fixed composition, as shown in figure \ref{fig:Angell V,eps}, the fragility decreases as $\epsilon_b$ increases. For liquids with weaker bonds ($\epsilon_b/\epsilon_{LJ}=5$), the fragility at constant pressure shows a non-monotonic dependence on network glass former composition. Several network materials, including alkali germanates \cite{shelby_viscosity_1974}, chalcogenides \cite{bohmer_correlations_1992}, and aluminosilicates \cite{bechgaard_fragility_2017} have been found to display such a non-monotonic dependence of the fragility on chemical composition. It is not clear whether all these chemical trends can be quantitatively explained within the present simple model. Another possibility suggested to explain the germanate anomaly \cite{welch_topological_2020} is the formation of 5- and 6- coordinated germanium as well as small rings of atoms. Such structures are not included in the present model, but could in principle be included specifically using the self-consistent phonon method and the RFOT theory methods. The non-monotonic variation in fragility due to composition change in chalcogenide glasses has been attributed to rigidity percolation effects \cite{phillips_topology_1979, thorpe_continuous_1983, tatsumisago_fragility_1990}, since the observed fragility reaches a global minimum at an apparent network coordination of 2.4, which would also be the rigidity percolation threshold. While rigidity theory has many interesting predictions \cite{micoulaut_concepts_2016}, it suffers by being rather vague at relating quantities directly to the forces. The temperature dependent rigidity theory resembles the present approach in that the fragility of network liquids in both cases is determined based on a calculation of the configurational entropy. The current work proposes an alternative calculation scheme for determining the configurational entropy as a function of composition and temperature. This calculation scheme is set up in the context of the RFOT theory which naturally implies the Adam-Gibbs relation between configurational entropy and viscosity. The actual Adam-Gibbs analysis \cite{adam_temperature_1965} is in fact quite different from the RFOT theory and predicts too small of a cooperative region. We believe there is value in using a starting point where the Adam-Gibbs empirical relation has firm theoretical footing and building the model up from there. The present calculations exhibit no notable features at this connectivity ($x=0.4$). Despite this, the current model is phenomenologically rich, and to investigate it more thoroughly, we here focus on the behavior of a liquid near $T_g$ as a function of $x$, $\epsilon_b$, and $P$ or $\rho$.

Notably, in both figure \ref{fig:mvvsx} and \ref{fig:mpvsx}, we see that the fragility can vary non-monotonically with respect to $x$ when the bonds are sufficiently weak. There is no non-monotonicity when bonds are very strong, consistent with what is seen for the silicates. The reason for this behavior becomes clear when we write the fragility explicitly in terms of the variables of our model. Specifically, fragility varies from two different effects
\begin{eqnarray}
    m&=&\frac{\partial \log \eta}{\partial (T_g/T)}\nonumber\\
    &=&\log e\frac{\partial (32/s_c)}{\partial (T_g/T)}\nonumber\\
    &=&-22T_g\left[\frac{\partial s_c}{\partial n_b}\frac{\partial n_b}{\partial T}+\frac{\partial s_c}{\partial \rho_{HS}^\mathrm{eff}}\frac{\partial \rho_{HS}^\mathrm{eff}}{\partial T}\right]\nonumber\\
    &=&m^{(n_b)}+m^{(\rho)}\label{eq:m deriv}
\end{eqnarray}
We see that if, $\epsilon_b$ is very large, $n_b$ stays relatively fixed as temperature is varied. As a result, $m^{(n_b)}$ is typically negligible when bond energies are large while $m^{(\rho)}$ monotonically decreases with increasing bond number per particle. If $\epsilon_b$ is small however, the number of bonds varies with termperature so $\frac{\partial n_b}{\partial T}$ is peaked at $n_b=2(1-x)$. For smaller $\epsilon_b$, we see that $m^{(n_b)}$ becomes significant, giving rise to the nonmonotonic behavior of $m$ with respect to $x$. As discussed earlier, some network materials do in fact display such a non-monotonic dependence of fragility on chemical composition. We see that one explanation may be that the bonds in these materials are relatively weak compared to those of silicates.

\section{Polyamorphism}
Polyamorphism refers to a scenario in which two distinct relatively dense liquid phases are manifested. In mixtures, this can occur trivially if the liquid phases of each pure component are immiscible. Single-component substances can sometimes form multiple distinct liquid phases if the pair potential has multiple length scales \cite{franzese_generic_2001} or if there are distinct structural motifs which compete in a frustrated manner \cite{tanaka_general_2000}. Changes of bonding provide a common mechanism for polyamorphism in real materials, such as silica \cite{mukherjee_direct_2001}, germania \cite{smith_equation_1995}, and water \cite{mishima_apparently_1985}. Often, the equilibrium bond distance will be greater than the diameter of the Mayer f-bond for the repulsive forces allowing the liquid to expand upon bonding. A particle with intact bonds typically has a locally tetrahedral structure (or some other geometry based on the number of bonds), while a particle without bonds will have a tendency to form close packed structures similar to a hard sphere liquid. These differences in length scale and liquid structure at the microscopic level lead to two distinct yet relatively dense phases at the macroscopic level. In one phase, the bonds are mostly intact and the liquid has a lower density to accommodate the longer bond distances. In the other phase, the bonds have become mostly broken and the liquid has a higher density as it is able to access high density more closely packed structures. These phases are often called the low density liquid (LDL) and high density liquid (HDL).

Even under conditions away from the phase separation, polyamorphism is often associated with a maximum in the density as a function of temperature \cite{angell_density_1976}, most well known in the case of water. A density maximum occurs even above temperatures where there are distinct liquid phases, and thus a density maximum can be used to suggest the presence of an underlying liquid-liquid phase separation. The nonmonotonic variation of density with temperature in turn leads to other nonmonotonic properties, such as the heat capacity in water \cite{angell_heat_1982}. Some alkali silicates also feature a density maximum as a function of composition \cite{tischendorf_density_1998}. It is therefore likely in our view that nonmonotonicities as a function of composition may also be a manifestation of an underlying phase transition. A nonmonotonic fragility (such as is seen in germanates) could then be explained directly by an underlying polyamorphic phase transition, due to the relation between $m$ and $\Delta c_p$ predicted by the RFOT theory. Indeed, germanate glasses across a variety of compositions do exhibit phase separation \cite{topping_properties_1974, alderman_germanate_2017}. In these cases, the phases seem to be a germanate rich ``bonded'' phase and a metal oxide rich ``non-bonded'' phase. We note that a density maximum arising from differences in packing bonded and non-bonded configurations can be distinguished from the nonmonotonicity occurring in the present model, which directly arises from the competition between changes in bond number and changes in density as a function of temperature. We feel both effects are present in systems with sufficiently weak bonds and density changes correlated with a liquid-liquid phase transition.

\section{Further discussion/conclusion}
Network materials are an important class of glass forming liquids. The fact that the  viscosity of many highly bonded liquids follows a nearly Arrhenius temperature dependence has lead to some debate over the physical interpretation of the activation energy in these materials suggesting even that transport was limited by bond breaking events. The RFOT theory suggests to a good approximation \cite{xia_fragilities_2000, rabochiy_microscopic_2013} that the free energy barrier to rearrangement can be computed based on the configurational entropy \cite{kirkpatrick_scaling_1989}, and since highly bonded liquids have a low configurational heat capacity, they will also have a roughly constant activation energy. Hall and Wolynes \cite{hall_microscopic_2003} explained using the self consistent phonon method that the $\Delta c_p$ of a glass former decreases with increasing number of bonds. We see this picture explains many of the patterns seen when varying the composition, density, and pressure of network fluids.

We have shown how to include the temperature dependence on the effective hard sphere density $\rho_{HS}^\mathrm{eff}$ and the number of bonds per particle $n_b$. In doing so, we have explained many experimental trends, including how $T_g$ increases with the introduction of bonds, while the fragility increases when a nonbonding impurity is added to a network former, as long as the bonds were sufficiently strong. Materials with weaker bonds are predicted to have a non-monotonic fragility as a function of bond number, as is sometimes seen experimentally. Several features of the network liquids near their glass transitions however suggest that fragility may also be influenced by an impending liquid-liquid transitions between high and low density amorphous forms which is an issue to which we hope to return in the future.

\begin{acknowledgments}
This research was supported by the Center for Theoretical Biological Physics sponsored by the NSF (Grants PHY-2019745). P.G.W. is also supported by the D.R. Bullard-Welch Chair at Rice University (Grant C-0016).
\end{acknowledgments}

\section*{Data Availability Statement}

Data sharing is not applicable to this article as no new data were created or analyzed in this study.
\section*{Conflict of Interest Statement}
The authors have no conflicts to disclose.

\bibliography{references}

\end{document}